\documentclass{article}

\usepackage[framemethod=tikz]{mdframed}
\usepackage{microtype}
\mdfdefinestyle{mystyle}{%
	nobreak=true,
	backgroundcolor=gray!20,
	innerleftmargin=10,
	innerrightmargin=10,
	outerlinewidth=3pt,
	topline=false,
	rightline=false,
    leftline=false,
	bottomline=false,
	skipabove=\topsep,
	skipbelow=\topsep
}

\usepackage{hyperref}

\hypersetup{hidelinks}
\usepackage{authblk}

\newcommand{\JournalTitle}[1]{{#1}}
\newcommand{\ket}[1]{|#1\rangle}

\title{Quantum Machine Learning}

\author[1,2,*]{Jacob Biamonte}
\author[3]{Peter Wittek}
\author[4]{Nicola Pancotti}
\author[5]{Patrick Rebentrost}
\author[6]{Nathan Wiebe}
\author[7]{Seth Lloyd}

\affil[*]{jacob.biamonte@qubit.org}

\affil[1]{Quantum Software Initiative, Skolkovo Institute of Science and Technology, Skoltech Building 3, Moscow 143026, Russia}
\affil[2]{Institute for Quantum Computing, University of Waterloo, Waterloo, N2L 3G1 Ontario, Canada}
\affil[3]{ICFO-The Institute of Photonic Sciences, Castelldefels (Barcelona), 08860 Spain}
\affil[4]{Max Planck Institute of Quantum Optics \\ Hans-Kopfermannstr. 1, D-85748 Garching, Germany}
\affil[5]{Massachusetts Institute of Technology, Research Laboratory of Electronics, Cambridge, MA 02139}
\affil[6]{Station Q Quantum Architectures and Computation Group, Microsoft Research, Redmond WA 98052}
\affil[7]{Massachusetts Institute of Technology, Department of Mechanical Engineering, Cambridge MA 02139 USA}

\begin{document}
\maketitle
\begin{abstract}	    
Fuelled by increasing computer power and algorithmic advances, machine learning techniques have become powerful tools for finding patterns in data. Since quantum systems produce counter-intuitive patterns believed not to be efficiently produced by classical systems, it is reasonable to postulate that quantum computers may outperform classical computers on machine learning tasks. The field of quantum machine learning explores how to devise and implement concrete quantum software that offers such advantages. Recent work has made clear that the hardware and software challenges are still considerable but has also opened paths towards solutions.
\end{abstract}

\section*{Introduction}

Long before they possessed computers, human beings strove to find patterns in data. Ptolemy fit observations of the motions of the stars to a geocentric model of the cosmos, with complex epicycles to explain retrograde motions of the planets. In the 16th century, Kepler analyzed the data of Copernicus and Brahe to reveal a previously hidden pattern: planets move in ellipses with the sun at one focus of the ellipse. The analysis of astronomical data to reveal such patterns gave rise to mathematical techniques such as methods for solving linear equations (Newton-Gauss), learning optima via gradient descent (Newton), polynomial interpolation (Lagrange), and least-squares fitting (Laplace). The nineteenth and early twentieth century gave rise to a broad range of mathematical methods for analyzing data to learn the patterns that it contained.  

The construction of digital computers in the mid 20th century allowed the automation of data analysis techniques. Over the past half century, the rapid progression of computer power has allowed the implementation of linear algebraic data analysis techniques such as regression and principal component analysis, and lead to more complex learning methods such as support vector machines. Over the same time frame, the development and rapid advance of digital computers spawned novel machine learning methods. Artificial neural networks such as perceptrons were implemented in the 1950s~\cite{rosenblatt1958perceptron}, as soon as computers had the power to realize them. Deep learning built on neural networks such as Hopfield networks and Boltzmann machines, and training methods such as back propagation, were introduced and implemented in the 1960s to 1990s~\cite{lecun2015deep}. In the past decade, particularly in the past five years, the combination of powerful computers and special-purpose information processors capable of implementing deep networks with billions of weights~\cite{le2013building}, together with their application to very large data sets, has revealed that such deep learning networks are capable of learning complex and subtle patterns in data. 

Quantum mechanics is well-known to generate counter intuitive patterns in data. Classical machine learning methods such as deep neural networks frequently have the feature that they can both recognize statistical patterns in data, and produce data that possess the same statistical patterns: they recognize the patterns that they produce. This observation suggests the following hope. If small quantum information processors can produce statistical patterns that are computationally difficult to be produced by a classical computer, then perhaps they can also recognize patterns that are equally difficult to recognize classically.

The realization of this hope depends on whether efficient quantum algorithms can be found for machine learning. A quantum algorithm is a set of instructions solving a problem, for example finding out whether two graphs are isomorphic, that can be performed on a quantum computer. Quantum machine learning software makes use of quantum algorithms as part of a larger implementation. Analysing the steps that quantum algorithms prescribe, it becomes clear that they have the potential to outperform classical algorithms for specific problems. This potential is known as quantum speedup.

The notion of a quantum speedup depends on whether one takes a formal computer science perspective---which demands mathematical proofs---or a perspective based on what can be done with realistic, finite-size devices---which requires solid statistical evidence of a scaling advantage over some finite range of problem sizes.  For the case of quantum machine learning, the best possible performance of classical algorithms isn't always known. This is similar to the case of Shor's polynomial-time quantum algorithm for integer factorization: no subexponetial-time classical algorithm has been found, but the possibility is not provably ruled out.  

Determination of a scaling advantage contrasting quantum and classical machine learning would rely on the existence of a quantum computer and is a so called, benchmarking problem.  Such advantages could include improved classification accuracy and sampling of classically inaccessible systems.  Accordingly,  quantum speedups in machine learning are currently characterized using idealized measures from complexity theory: query complexity and gate complexity (see Box 1). Query complexity measures the number of queries to the information source for the classical or quantum algorithm. A quantum speedup results if the number of queries needed to solve a problem is lower for the quantum- than for the classical algorithm. To determine the gate complexity, the number of elementary quantum operations, or gates, required to obtain the desired result are counted.
 
Query and gate complexity are idealized models which quantify the necessary resources to solve a problem class. Without knowing how to map this idealization to reality, not much can be said about the necessary resource scaling in a real-world scenario. Therefore, the required resources of classical machine learning algorithms are mostly quantified by numerical experimentation. The resource requirements of quantum machine learning algorithms are likely to be similarly difficult to quantify in practice. The analysis of their practical feasibility is a central subject of this review.
  
As will be seen throughout the review, there are quantum algorithms for machine learning that exhibit quantum speedups~\cite{2015ConPh..56..172S,wittek2014qml,Adcock2015Advances,arunchalam2017survey}. For example, the quantum basic linear algebra subroutines (BLAS)---Fourier transforms, finding eigenvectors and eigenvalues, solving linear equations---exhibit exponential quantum speedups over their best known classical counterparts~\cite{2009PhRvL.103o0502H,2012PhRvL.109e0505W,childs2015quantum}. This quantum BLAS (qBLAS) translates into quantum speedups for a variety of data analysis and machine learning algorithms including linear algebra, least-squares fitting, gradient descent, Newton's method, principal component analysis, linear, semidefinite, and quadratic programming, topological analysis, and support vector machines~\cite{2012PhRvL.109e0505W,2014NatPh..10..631L,kimmel2016hamiltonian,2014PhRvL.113m0503R,lloyd2014quantumalgorithms,dridi2015homology,rebentrost2016quantumsingular,schuld2016prediction,brandao2016quantum,rebentrost2016quantum}. At the same time, special purpose quantum information processors such as quantum annealers and programmable quantum optical arrays are well-matched to deep learning architectures~\cite{Wiebe2014Quantum, adachi2015application, amin2016quantum}. While it is not completely clear yet to which extent this potential is born out in reality, there are reasons to be optimistic that quantum computers can recognize patterns in data that surpass the reach of classical computers.

We organize this review as follows. The machines that learn can be either classical~\cite{sasaki2001quantum, 2010PhRvA..81c2324B, bisio2011quantumlearning, 2012arXiv1208.0663S, sentis2014quantum, paparo2014quantum, dunjko2015quantum-enhanced, dunjko2016quantumenhanced, sentis2016quantumchange, 2014PhRvX...4d1012F} or quantum~\cite{2009PhRvL.103o0502H, 2013PhRvL.110y0504C, 2012PhRvL.109e0505W, lloyd2013quantum, 2014PhRvL.113m0503R, 2014arXiv1401.2142W, 2014NatPh..10..631L, lau2016quantum}. The data they analyze can be either classical or quantum states produced by quantum sensing or measuring apparatus~\cite{aimeur2006machine,dunjko2016quantumenhanced}. We briefly discuss conventional machine learning---the use of classical computers to find patterns in classical data. We then turn to quantum machine learning, where the data that the quantum computer analyzes can be either classical data, which ends up encoded as quantum states, or quantum data. Finally, we discuss briefly the problem of using classical machine learning techniques to find patterns in quantum dynamics.  

\begin{mdframed}[style=mystyle,frametitle=Box 1. Quantum Speedups]

\small{
Quantum computers use effects such as quantum coherence and entanglement to process information in ways that classical computers can not.  The past two decades have seen steady advances in constructing more powerful quantum computers. A quantum algorithm is a step-wise procedure performed on a quantum computer to solve a problem, such as searching a database.  Quantum machine learning software makes use of quantum algorithms to process information.  

Quantum algorithms can in principal outperform the best known classical algorithms when solving certain problems. This is known as a quantum speedup \cite{2014Sci...345..420R}. The question addressed in this review is how quantum computers and special-purpose information processors such as quantum annealers could be used to perform quantum machine learning.

For example, quantum computers can search an unsorted database with $N$ entries in time proportional to  $\sqrt{N}$---that is $O(\sqrt{N})$---where a classical computer given blackbox access to the same database takes time proportional to $N$: the quantum computer exhibits a square root speed up over the classical computer.  Similarly, quantum computers can perform Fourier transforms over $N$ data points, invert sparse $N\times N$ matrices, and find their eigenvalues and eigenvectors in time proportional to a polynomial in $\log_2 N$, where the best known algorithms for classical computers take time proportional to $N \log_2 N$: the quantum computer exhibits an exponential speed up over the best classical computer algorithms.  The following table summarizes the techniques used to achieve speedups for various quantum-enhanced machine learning subroutines.  

In the above table, speedups are taken with respect to their classical counterpart(s)---hence, 
$O(\sqrt{N})$ means quadratic speedup and $O(\log(N))$ means exponential relative to their classical counterpart.  $^{(*)}$ denotes important caveats that can limit applicability of method~\cite{aaronson2015read} and AA denotes amplitude amplification. \\

{\footnotesize
\centering
\resizebox{\columnwidth}{!}{
\begin{tabular}{|c|c|c|c|c|c|}
\hline
Method                          & Speedup                               &AA                     & HHL                   &Adiabatic              & QRAM\\
\hline
Bayesian Inference~\cite{low2014quantum,wiebe2015can}      &$O(\sqrt{N})$                             &Y                              & Y                     &N                                      & N\\
Online Perceptron~\cite{wiebe2016quantumperceptron}         & $O(\sqrt{N})$                            & Y                     &N                      &N                                      & optional\\
Least squares fitting~\cite{2012PhRvL.109e0505W}& $O(\log{N}^{(*)})$           & Y                     &Y                              &N                                      &Y\\
Classical BM~\cite{Wiebe2014Quantum}&$O(\sqrt{N})$                             &Y/N                    &optional/N     &N/Y                            &optional\\
Quantum BM~\cite{amin2016quantum,kieferova2016tomography}&$O(\log{N}^{(*)})$               &optional/N     &N                      &N/Y                            &N\\
Quantum PCA~\cite{2014NatPh..10..631L}&$O(\log{N}^{(*)})$              &N      &Y                      &N                              &optional\\
Quantum SVM~\cite{2014PhRvL.113m0503R}&$O(\log{N}^{(*)})$              &N      &Y                      &N                              &Y\\
Quantum reinforcement learning~\cite{dunjko2016quantumenhanced}&$O(\sqrt{N})$                          &Y                      &N      &N                              &N\\
\hline
\end{tabular}
}\\
}
}

\end{mdframed}

\section*{Classical machine learning}

Classical machine learning and data analysis can be divided into several categories. First, computers can be used to perform `classic' data analysis methods such as least squares regression, polynomical interpolation, and data analysis. Machine learning protocols can be supervised or unsupervised. In supervised learning, the training data is divided into labeled categories, such as samples of handwritten digits together with the actual number the handwritten digit is supposed to represent, and the job of the machine is to learn how to assign labels to data outside the training set. In unsupervised learning, the training set is unlabeled: the goal of the machine is to find the natural categories into which the training data falls (e.g., different types of photos on the internet) and then to categorize data outside of the training set. Finally, there are machine learning tasks, such as playing Go, that involve combinations of supervised and unsupervised learning, together with training sets that may be generated by the machine itself. 

\section*{Linear-algebra based quantum machine learning}

A wide variety of data analysis and machine learning protocols operate by performing matrix operations on vectors in a high dimensional vector space. But quantum mechanics is all about matrix operations on vectors in high dimensional vector spaces.  

The key ingredient behind these methods is that the quantum state of $n$ quantum bits or qubits is a vector in a $2^n$-dimensional complex vector space; quantum logic operations or measurements performed on qubits multiplies the corresponding state vector by $2^n \times 2^n$ matrices. By building up such matrix transformations, quantum computers have been shown to perform common linear algebraic operations such as Fourier transforms~\cite{1995quant.ph..8027S}, finding eigenvectors and eigenvalues~\cite{nielsen2000quantum}, and solving linear sets of equations over $2^n$-dimensional vector spaces in time polynomial in $n$, exponentially faster than their best known classical counterparts~\cite{2009PhRvL.103o0502H}. This latter is commonly referred to as the HHL algorithm for the authors of the paper see Box 2).  The original variant assumed a well-conditioned matrix that is sparse.  Sparsity is unlikely in data science, but later improvements relaxed this assumption to include low-rank matrices as well~\cite{2013PhRvL.110y0504C,childs2015quantum,wossnig2017quantum}. Going past HHL, here we survey several quantum algorithms which appear as subroutines when linear algebra techniques are employed in quantum machine learning software.  

\subsection*{Quantum principal component analysis}

For example, consider principal component analysis (PCA). Suppose that one's data is presented in the form of vectors $\vec v_j$ in a $d$ dimensional vector space. For example, $\vec v_j$ could be the vector of changes in prices of all stocks in the stock market from time $t_j$ to time $t_{j+1}$. The covariance matrix of the data is $C = \sum_j \vec v_j \vec v_j^T$, where $T$ denotes the transpose operation: the covariance matrix summarizes the correlations between the different components of the data, e.g., correlations between changes in the prices of different stocks. In its simplest form, principal component analysis operates by diagonalizing the covariance matrix: $C = \sum_k e_k \vec c_k \vec c_k^\dagger$, where the $\vec c_k$ are the eigenvectors of $C$, and $e_k$ are the corresponding eigenvalues. (Because $C$ is symmetric, the eigenvectors $\vec c_k$ form an orthonormal set.) If only a few of the eigenvalues $c_k$ are large, and the remainder are small or zero, then the eigenvectors corresponding to those eigenvalues are called the principal components of $C$. Each principal component represents an underlying common trend or form of correlation in the data, and decomposing a data vectors $\vec v$ in terms of principal components, $\vec v = \sum_k v_k \vec c_k$ allows one both to compress the representation of the data, and to predict future behavior. Classical algorithms for performing PCA scale as $O(d^2)$ in terms of computational complexity and query complexity.

For quantum principal component analysis of classical data (qPCA~\cite{2014NatPh..10..631L}), we choose a data vector $\vec v_j$ at random, and use a quantum random access memory (qRAM~\cite{giovannetti2008quantum}) to map that vector into a quantum state: $\vec v_j \rightarrow |v_j\rangle$. The quantum state that summarizes the vector has $\log d$ qubits, and the operation of the qRAM requires $O(d)$ operations divided over $O(\log d)$ steps that can be performed in parallel. Because $\vec v_j$ was chosen at random, the resulting quantum state has a density matrix $\rho = (1/N) \sum_j |v_j\rangle \langle v_j|$, where $N$ is the number of data vectors. Comparing with the covariance matrix $C$ for the classical data we see that the density matrix for the quantum version of the data \emph{is} the covariance matrix, up to an overall factor. Repeatedly sampling the data, and using a trick called density matrix exponentiation~\cite{lloyd1996universal} combined with the quantum phase estimation algorithm~\cite{nielsen2000quantum}, which finds eigenvectors and eigenvalues of matrices, allowing one to take the quantum version of any data vector $|\vec v\rangle$ and to decompose it into the principal components $|c_k\rangle$, revealing the eigenvalue of $C$ at the same time:$|v\rangle \rightarrow \sum_k v_k |c_k\rangle|\tilde e_k\rangle$. The properties of the principal components of $C$ can then be probed by making measurements on the quantum representation of the eigenvectors of $C$. The quantum algorithm scales as $O\big( (\log d)^2 \big)$ in both computational complexity and query complexity. That is, quantum PCA is exponentially more efficient than classical PCA.

\subsection*{Quantum support vector machines and kernel methods}

The simplest examples of supervised ML algorithms are linear support vector machines and perceptrons. These methods seek to find an optimal separating hyperplane between two classes of data in a data set such that, with high probability, all training examples of one class are only found on one side of the hyperplane. The most robust classifier for the data is given when the margin between the hyperplane and the data is maximized. Here the ``weights'' learned in the training are the parameters of the hyperplane. One of the greatest powers of the SVM lies in its generalization to nonlinear hyper-surfaces via kernel functions~\cite{vapnik1995nature}. Such classifiers have found great success in image segmentation as well as in the biological sciences. 

As its classical counterpart, the quantum support vector machine is a paradigmatic example of a quantum machine learning algorithm~\cite{2014PhRvL.113m0503R}. A first quantum support vector machine was discussed in the early 2000s~\cite{anguita2003quantum}, using a variant of Grover's search for function minimization \cite{durr1996quantum}. Finding $s$ support vectors out of $N$ vectors consequently takes $\sqrt{N/s}$ iterations. Recently a least-squares quantum support vector machine was developed that harnesses the full power of the qBLAS subroutines. The data input can come from various sources, such as from qRAM accessing classical data or a quantum subroutine preparing quantum states. Once the data are made available to the quantum computing device, they are processed with quantum phase estimation and matrix inversion (the HHL algorithm). All the operations required to construct the optimal separating hyperplane and to test whether a vector lies on one side or the other can in principle be performed in time ${\rm poly}(\log N)$ where $N$ is the dimension of the matrix required to prepare a quantum version of the hyperplane vector. Polynomial\cite{2014PhRvL.113m0503R} and radial basis function kernels \cite{chatterjee2016generalized} are discussed, as well as another kernel-based method called Gaussian process regression~\cite{2015arXiv151203929Z}.  This approach to quantum support machines has been experimentally demonstrated in a nuclear magnetic resonance testbed for a hand-written digit recognition task \cite{2014arXiv1410.1054Z}.

\begin{mdframed}[style=mystyle,frametitle=Box 2. HHL Algorithm]
\footnotesize{The HHL algorithm for inverting systems of equations is a fundamental, and easy to understand subroutine, underpinning many quantum machine learning algorithms. The algorithm seeks to solve $A \vec x = \vec b$ using a quantum computer. HHL quantizes the problem by expressing the vector $\vec b \in C^N$ as a quantum state $\ket{b}$ over $\log_2 N$ qubits, and the vector $\vec x$ as a quantum state $\ket{x}$.  The matrix $A$ can be assumed to be Hermitian without loss of generality because the space can always be expanded to make this true.
The equation $A\ket{x} = \ket{b}$ can then be solved by multiplying both sides of the equation by $A^{-1}$.  The Harrow, Hassidim and Lloyd algorithm then allows
one to construct the quantum state proportional $A^{-1} \ket{b}$, where $A^{-1}$ is the inverse of $A$.   More generally, when $A$ is not 
square or has zero eigenvalues, the algorithm can be used to find the state
$\ket{x}$ that minimizes $| A\ket{x} - \ket{b}|$~\cite{2012PhRvL.109e0505W}.

The algorithm works as follows.  Assume $\ket{b} = \sum_n b_n \ket{E_n}$ where $\ket{E_n}$ is an eigenvector of $A$ with eigenvalue $\lambda_n\ge \Lambda$.  By applying phase estimation under $A$ to compute $\lambda_n$  and by rotating an ancillary qubit through an angle of $\arcsin(\Lambda/\lambda_n)$ and then uncomputing the phase-estimation we obtain 
$$
 \sum_n b_n \ket{E_n} \left( \frac{\Lambda}{\lambda_n} \ket{1} + \sqrt{1-\frac{\Lambda^2}{\lambda_n^2}}\ket{0}\right).
$$
Then if the ancillary qubit is measured and if $1$ is observed then each eigenstate is divided through by $\lambda_n$, which affects the inverse.  The number of times that the state preparation circuit needs to be applied to succeed, after applying amplitude amplification, is $O(\|A\|/\Lambda)$ which is the condition number for the matrix.

The HHL algorithm takes $O( (\log N)^2 )$ quantum steps to output $\ket{x}$, compared with $O(N \log N)$ steps required to find $\vec x$
using the best known method on a classical computer.   

There are several important caveats to the HHL algorithm.  First, finding the full answer
$\vec x$ from the quantum state $\ket{x}$ requires $O(N)$ repetitions to reconstruct
the $N$ components of $\vec x$. Generalizations to HHL such as least squares fitting side step this problem by allowing the output to be much smaller dimensional than the input.  In general, however, HHL can only provide features of the data such as moments of the solution vector or its expectation value $\vec x^\dagger B \vec x$ over other sparse matrices $B$.
The second caveat is that the input vector $\ket{b}$ needs to be prepared, either on a quantum computer or using qRAM, which may be expensive.
The third caveat is that the matrix must be well conditioned and $e^{-iA}$ must be efficiently simulatable.  Finally, although the HHL algorithms scales as $O( (\log N)^2 )$
current estimates of the cost of the algorithm for practical problems are prohibitive~\cite{scherer2017concrete}, which underlines the importance of investigating further improvements such as~\cite{childs2015quantum}.
In general, the promise of exponential speedups for linear systems should be tempered with the realization that they only apply to certain problems.
}

\end{mdframed}

\begin{figure*}[!t]
\centering
\includegraphics[width=1.0\textwidth]{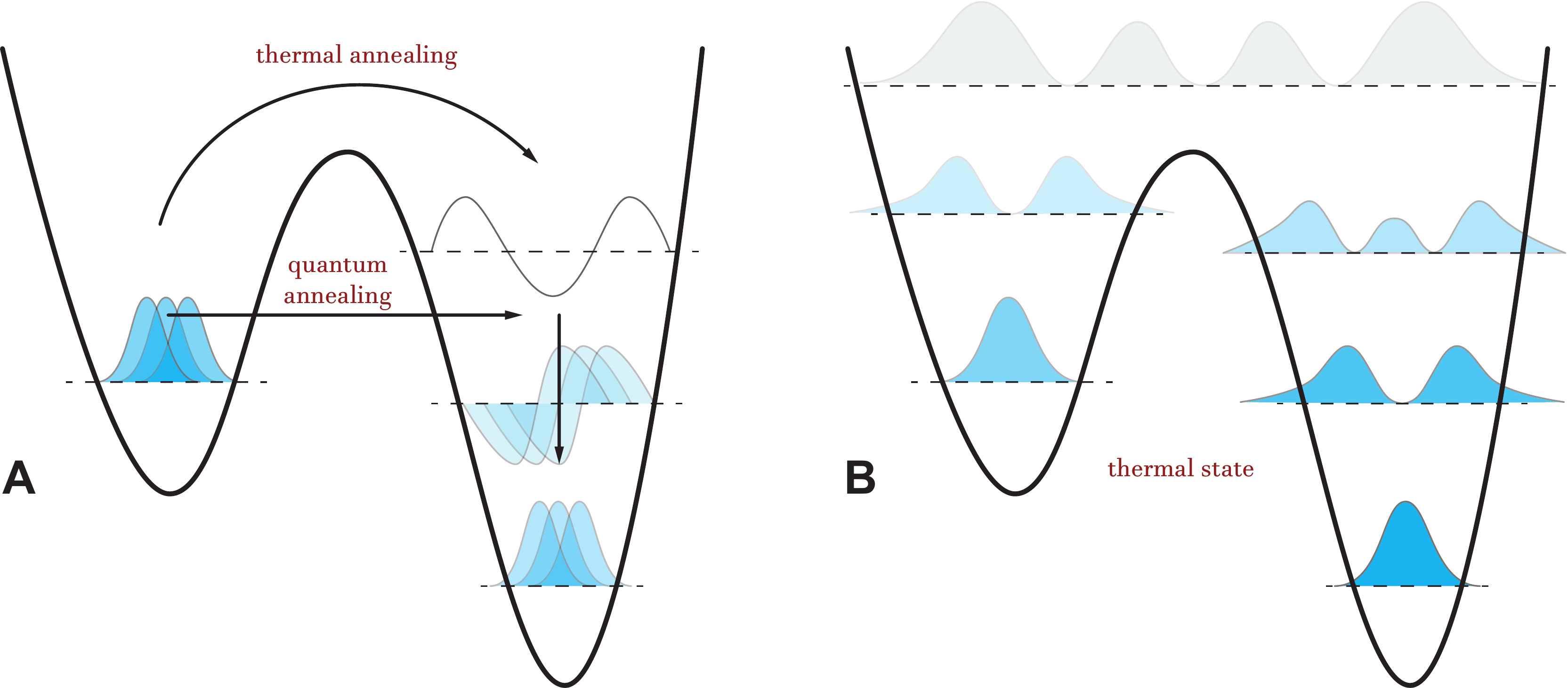}
\caption{{\bf Quantum tunneling vs thermalization.} A quantum state tunnels when approaching a resonance point before decoherence induces thermalization.  {\bf A}.~A quantum state must traverse a local minimum in thermal annealing whereas a coherent quantum state can tunnel when brought close to resonance.  {\bf B}.~Coherent effects decay through interaction with an environment, resulting in a probability distribution in occupancy of a systems energy levels following a Gibbs distribution.  }
\label{fig:aqc-gibbs}
\end{figure*}

\section*{qBLAS-based optimization}

A wide variety of data analysis and machine learning techniques involve
optimization.  Of increasing interest is the use of D-Wave processors to solve combinatorial optimization problems by means of quantum annealing.  Some optimization problems can also be formulated as a single shot solution of a linear system, for example the optimization of a quadratic function subject to equality constraints, a subset of quadratic programming problems. If the matrices involved are sparse or low rank, such problems can be solved in time ${\rm poly}(\log d)$, where $d$ is the system dimension via the HHL matrix inversion algorithm, yielding an exponential speedup over classical algorithms, which run in time ${\rm poly}(d)$.

Most methods in ML require iterative optimization of their performance. As an example, inequality constraints are often handled via penalty functions \cite{2012EL.....9957004W} and variations of gradient descent or Newton's method. A modification of the quantum PCA method implements iterative gradient descent and Newton's methods for polynomial optimization, again providing an exponential speedup over classical methods.\cite{rebentrost2016quantum} Multiple copies of the present solution, encoded in a quantum state, are used to improve that solution at each step. Brandao and Svore provide a quantum version of semi-definite programming, that holds out the possibility of super-polynomial speed-ups~\cite{brandao2016quantum}. The quantum approximate optimization algorithm (QAOA, or QAO algorithm)~\cite{farhi2014quantum} provides a unique approach to optimization based on alternating qubit rotations with the application of the problem's penalty function.

\section*{Reading classical data into quantum machines}

Classical data must be inputted before being processed on a quantum computer.  Often called, `the input problem', this is often done with little overhead but does present a serious bottle neck for certain algorithms.  Likewise, the `output problem' is faced when reading out data after being processed on a quantum device.  Like the input problem, the output problem often brings a significant operational slowdown.  

Specially, if we consider applying HHL, least squares fitting, qPCA, quantum support vector machines, and related approaches need to classical data, the procedure begins with first loading considerable amounts of data into a quantum system which can require exponential time~\cite{aaronson2015read}.  This can be addressed in principle using qRAM but the cost of doing so may be prohibitive for big data problems~\cite{arunachalam2015robustness}.  Apart from combinatorial optimization based approaches, the only known linear-algebra based quantum machine learning algorithm that does not rely on large-scale qRAM is the quantum algorithm for performing topological analysis of data (persistent homology) \cite{lloyd2016quantum}. With the notable exceptions of least squares fitting and quantum support vector machines, linear algebra based algorithms also can suffer from the output problem since classical quantities that are sought after such as the solution vector for HHL or the principal components for PCA are exponentially hard to estimate.  

Despite the potential for exponential quantum speed ups, without significant effort put into optimization, the circuit size and depth overhead can balloon
(to $\sim 10^{25}$ in one proposed realization of HHL \cite{scherer2015resource}).   Ongoing work is needed to optimize such algorithms, provide better cost estimates and ultimately to understand the sort of quantum computer that we would need to provide useful quantum alternatives to classical machine learning within this space.

\section*{Deep quantum learning}

Classical deep neural networks are highly effective tools for machine learning and are well suited to inspire the development of deep quantum learning methods. Special-purpose quantum information processors such as quantum annealers and programmable photonic circuits are well-suited for constructing deep quantum learning networks~\cite{denil2011toward,dumoulin2013challenges,adachi2015application}. The simplest deep neural network to quantize is the Boltzmann machine. The classical Boltzmann machine consists of bits with tunable interactions: the Boltzmann machine is trained by adjusting those interactions so that the thermal statistics of the bits, described by a Boltzmann-Gibbs distribution (see figure~\ref{fig:aqc-gibbs}B), reproduces the statistics of the data. To quantize the Boltzmann machine one simply takes the neural network and expresses it as a set of interacting quantum spins, corresponding to a tunable Ising model. Then by initializing the input neurons in the Boltzmann machines into a fixed state and allowing the system to thermalize, we can read out the output qubits to obtain an answer.
 
An essential feature of deep quantum learning is that it does not require a large, general purpose quantum computer. Quantum annealers are special purpose quantum information processors that are significantly easier to construct and to scale up than general purpose quantum computers (see figure~\ref{fig:aqc-gibbs}A). Quantum annealers are well-suited for implementing deep quantum learners, and are commercially available. The D-Wave quantum annealer is a tunable transverse Ising model that can be programmed to yield thermal states of classical, and certain quantum spin systems. The D-Wave device has been used to perform deep quantum learning protocols on more than a thousand spins~\cite{benedetti2015estimation}. Quantum Boltzmann machines~\cite{amin2016quantum} with more general tunable couplings, capable of implementing universal quantum logic, are currently in the design stage~\cite{2008PhRvA..78a2352B}. On chip silicon waveguides have been used to construct linear optical arrays with hundreds of tunable interferometers, and special purpose superconducting quantum information processors could be used to implement the QAO algorithm. 

There are several ways that quantum computers can give advantages here. First, quantum methods can make the system thermalize quadratically faster than its classical counterpart~\cite{temme2011quantum,yung2012quantum,Wiebe2014Quantum,chowdhury2016quantum}.  This can make accurate training of fully connected Boltzmann machines practical.

Second, quantum computers can accelerate Boltzmann training by providing improved ways of sampling. Because the neuron activation pattern in the Boltzmann machine is stochastic, many repetitions are needed to find success probabilities, and in turn the effect that changing a weight in the neural network has on the performance of the deep network. In training a quantum Boltzmann machine, by contrast, quantum coherence can quadratically reduce the number of samples needed to learn the performance. Furthermore, quantum access to the training data (i.e.~qRAM or a quantum blackbox subroutine) allows the machine to be trained using quadratically fewer accesses to the training data than classical methods requires: a quantum algorithm can train a deep neural network on a large training data set while only reading a minuscule number of training vectors~\cite{Wiebe2014Quantum}.
 
Quantum information processing provides new {fundamentally quantum} models for deep learning. For example, adding a transverse field to the simple Ising model quantum Boltzmann machine above yields a transverse Ising model, which can exhibit a variety of fundamental quantum effects such as entanglement~\cite{amin2016quantum,kieferova2016tomography}. Adding further quantum couplings transforms the quantum Boltzmann machine into a variety of quantum systems \cite{2008PhRvA..78a2352B,lloyd2016adiabatic}. Adding a tunable transverse interaction to a tunable Ising model is  known to be universal for full quantum computing~\cite{2008PhRvA..78a2352B}: with the proper weight assignments this model can execute any algorithm that a general purpose quantum computer can perform. Such universal deep quantum learners may recognize and classify patterns that classical computers can not.

Unlike classical Boltzmann machines, quantum Boltzmann machines output a quantum state. Thus deep quantum networks can learn to generate quantum states representative of a wide variety of systems.  This ability is absent from classical machine learning and allows it to act as a form of quantum associative memory~\cite{ventura2000quantum}.  Thus quantum Boltzmann training has applications beyond classifying quantum states and providing richer models for classical data.  

\begin{mdframed}[style=mystyle,frametitle=Box 3. Training Quantum Boltzmann Machines]
\small{In quantum Boltzmann machine training we wish to learn a set of Hamiltonian parameters ($w_j$) such that for a fixed set of $H_j$ we have that our input state $\rho_{\rm train}$ is well approximated by $\sigma=e^{-\sum_j w_j H_j}/{\rm Tr}( e^{-\sum_j w_j H_j})$~\cite{amin2016quantum,kieferova2016tomography}.   For all visible Boltzmann machines, the  quantum relative entropy $S(\rho_{\rm train}\|\sigma)= {\rm Tr} \left(\rho_{\rm train} \log(\rho_{\rm train}) -\rho_{\rm train} \log(\sigma)\right)$ is the most natural way to measure the quality of the approximation.  It is easy to see (assuming that the kernels of $\rho$ and $\sigma$ coincide) that the quantum relative entropy upper bounds the distance between the two states.  Thus, minimizing it minimizes the error in approximating the state.

While the relative entropy is an excellent measure of the distance between two states, it can be difficult to learn experimentally.  However, the gradient (i.e. direction of greatest change) of the relative entropy is easy to estimate~\cite{kieferova2016tomography}:
$$
\partial_{w_j} S(\rho_{\rm train}\|\sigma) = {\rm Tr}(\sigma H_j) – {\rm Tr}(\rho H_j).
$$
Given an experimental data set of expectation values for $\rho_{\rm train}$ and a quantum simulator for ${\rm Tr}(\sigma H_j)$ we can find the direction of greatest improvement in the quantum relative entropy.  Gradient descent then is used to update $\vec{w}$ via $\vec{w} \rightarrow \vec{w} - \eta \nabla S(\rho_{\rm train}\|\sigma)$ for $\eta>0$ .  Stoquastic Hamiltonians have the property that all off-diagonal matrix elements in the standard basis are real and non-positive (eqv.~non-negative).  No efficient classical analogue of this method is known in general for non-stoquastic $H$---see~\cite{2008PhRvA..78a2352B}.  

We show this protocol below for learning a random state formed from a uniform mixture of $4$ random states---random with respect to the unique and unitarily invariant Haar measure. Fewer than $10$ gradient steps (epochs) are needed to train it to approximately generate $\rho_{\rm train}$ using a complete set of Hamiltonian terms.
}
\\\\
\includegraphics[width=1.0\textwidth]{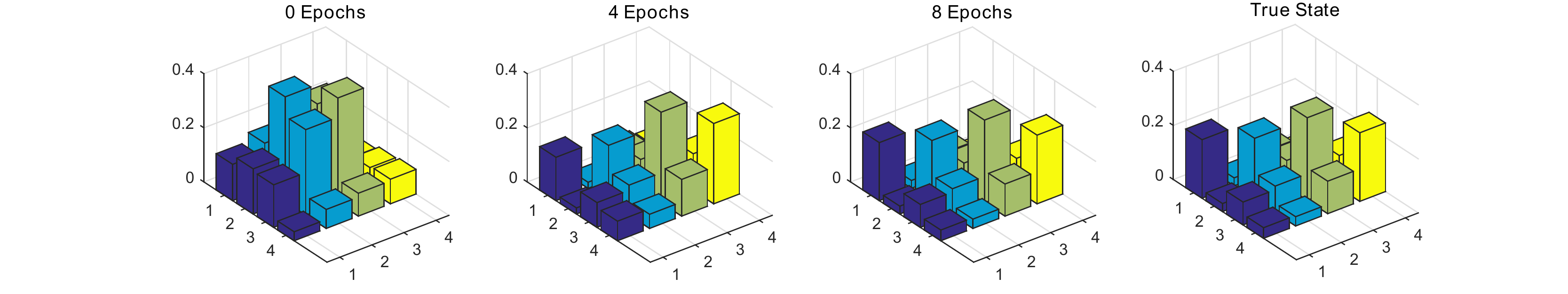}

\end{mdframed}
 
\section*{Quantum machine learning for quantum data}

Perhaps the most immediate application of quantum machine learning is to quantum data -- the actual states generated by quantum systems and processes. As described above, many quantum machine learning algorithms find patterns in classical data by mapping the data to quantum mechanical states, and then manipulating those states using basic quantum linear algebra subroutines. Those same quantum machine learning algorithms can be applied directly to the quantum states of light and of matter to reveal their underlying features and patterns. The resulting quantum modes of analysis are frequently much more efficient and more illuminating than the classical analysis of data taken from quantum systems. For example, given multiple copies of a system described by an $N\times N$ density matrix, quantum principal component analysis can be used to find its eigenvalues and to reveal the corresponding eigenvectors in time $O( (\log_2 N)^2)$, compared with $O(N^2)$ measurements needed for a classical device to perform tomography on density matrix, and
the $O(N^2)$ operations needed to perform the classical PCA.  Such quantum analysis of quantum data could profitably be performed on the relatively small quantum computers that are likely to be available over the next several years. 

A particularly powerful quantum data analysis technique is the use of quantum simulators to probe quantum dynamics. Quantum simulators are `quantum analog computers'---quantum systems whose dynamics can be programmed to match the dynamics of some desired quantum system. A quantum simulator can either be a special purpose device constructed to simulate a particular class of quantum systems, or a general purpose quantum computer. By connecting a trusted quantum simulator to an unknown system and tuning the model of the simulator to counteract the unknown dynamics, the dynamics of the unknown system can be efficiently learned using approximate Bayesian inference.\cite{granade2012robust,wiebe2014hamiltonian,wiebe2015quantum}
 This exponentially reduces the number of measurements needed to perform the simulation. Similarly, the universal quantum emulator algorithm~\cite{marvian2016universal}
allows one to reconstruct quantum dynamics and the quantum Boltzmann training algorithm of~\cite{kieferova2016tomography} allows states to be reconstructed, in time logarithmic in the dimension of the Hilbert space---exponentially faster than reconstructing the dynamics via classical tomography. 

In order to use a quantum computer to help characterize a quantum system~\cite{wiebe2014hamiltonian, wiebe2015quantum} or to accept input states for use in a quantum PCA algorithm then we must face the significant technical challenge of loading coherent input states.  Nonetheless, because such applications do not require QRAM and offer the potential for exponential speedups for device characterization~\cite{wiebe2014hamiltonian,wiebe2015quantum,amin2016quantum,kieferova2016tomography} they remain among the promising possibilities for near-term application of quantum machine learning.

\section*{Designing and controlling quantum systems}\label{control}

A major challenge in the development of quantum computation and information science involves tuning quantum gates to within the exacting requirements needed for quantum error correction. Heuristic search methods can help achieve this in a supervised learning scenario \cite{nvcontrol, zahedinejad2016designing}, for instance in the case of nearest-neighbor-coupled superconducting artificial atoms~\cite{zahedinejad2016designing} with gate fidelity above $99.9\%$ in the presence of noise, and hence reaching an accepted threshold for fault-tolerant quantum computing. A similar methodology has been successful in constructing a single-shot Toffoli gate, again reaching gate fidelity above $99.9\%$~\cite{zahedinejad2015high-fidelity}. Genetic algorithms have been employed to reduce digital and experimental errors in quantum gates~\cite{PhysRevA.64.023420}. They have been used to simulate CNOT gates by means of ancillary qubits and imperfect gates. Besides outperforming protocols for digital quantum simulations, it has been shown that genetic algorithms are also useful for suppressing experimental errors in gates~\cite{2015arXiv151200674L}.  Another approach used stochastic gradient descent and two body interactions to embed a Toffoli gate without time-dependent control using the natural dynamics of a quantum network~\cite{banchi2015quantumgate}. Dynamical decoupling sequences help protect quantum states from decoherence, which can be designed using recurrent neural networks~\cite{august2016using}.

Controlling a quantum system is just as important and complex. Learning methods have also seen ample success in developing control sequences to optimize adaptive quantum metrology, which is a key quantum building block in many quantum technologies. Genetic algorithms have been proposed for the control of quantum molecules to overcome the problem caused by changing in environmental parameters in an experiment~\cite{doi:10.1021/j100014a048}. Reinforcement learning algorithms using heuristic global optimization, like the one in designing circuits, have been widely successful, particularly in the presence of noise and decoherence, scaling well with the system size~\cite{hentschel2010machine,lovett2013differential,palittapongarnpim2016learning}. One can also exploit reinforcement learning in gate-based quantum systems. For instance, adaptive controllers based on intelligent agents for quantum information demonstrate adaptive calibration and compensation strategies to an external stray field of unknown magnitude in a fixed direction.

Classical machine learning is also a powerful tool to extract theoretical insights about quantum states. Neural networks have recently been deployed to study two central problems in condensed matter, namely phase of matter detection~\cite{Carrasquilla2017,Broecker} and ground state search~\cite{Carleo602}. They gathered ample success achieving better performances than established numerical tools. Theoretical physicists are now studying these models to analytically understand their descriptive power compared to traditional methods such as tensor networks. Interesting applications to exotic states of matter are already on the market, it has been shown that they can capture highly non trivial features from disordered or topologically ordered systems.

\section*{Perspectives on future work}

As shown in this review, small quantum computers and larger special purpose quantum simulators, annealers, etc., exhibit promising applications in machine learning and data analysis~\cite{Brunner2013, 2015PhRvL.114k0504C, hermans2015photonic, tezak2015coherent, lau2016quantum, neigovzen2009quantum, 2014arXiv1410.1054Z, PhysRevLett.98.023003, neven2009nips, denchev2012robust,karimi2012investigating,ogorman2014bayesian,denchev2015totally, adachi2015application,dridi2015homology,amin2016quantum,kerenidis2016quantum,alvarez-rodriguez2016quantum,wittek2017quantumenhanced}. The execution of these algorithms requires quantum hardware: can this promise be realized?

On the hardware side, there have been great strides in several enabling technologies. Small scale quantum computers with 50-100 qubits will be made widely available via quantum cloud computing (the `Qloud'). Special purpose quantum information processors such as quantum simulators, quantum annealers, integrated photonic chips, NV-diamond arrays, quantum random access memory, and made-to-order superconducting circuits will continue to advance in size and complexity. Quantum machine learning offers a suite of potential applications for small quantum computers~\cite{sasaki2001quantum, 2010PhRvA..81c2324B, bisio2011quantumlearning, 2012arXiv1208.0663S, sentis2014quantum, paparo2014quantum, dunjko2015quantum-enhanced, dunjko2016quantumenhanced, sentis2016quantumchange, lamata2017basic, schuld2017quantum, AW17} complemented and enhanced by special purpose quantum information processors~\cite{adachi2015application, amin2016quantum}, digital quantum processors~\cite{tiersch2014adaptive,zahedinejad2015designing, zahedinejad2015high-fidelity, banchi2015quantumgate, palittapongarnpim2016learning} and sensors~\cite{hentschel2010machine,lovett2013differential,palittapongarnpim2016controlling}

In particular, quantum annealers with $\sim 2000$ qubits have been built and operated, using integrated superconducting circuits that are in principle scalable.   The biggest challenges for quantum annealers to implement quantum machine learning algorithms include improving connectivity and implementing more general tunable couplings between qubits. Programmable quantum optic arrays with $\sim 100$ tunable interferometers have been constructed using integrated photonics in silicon, but loss represents an important challenge in scaling such circuits up.  A particularly important challenge for quantum machine learning is the construction of interface devices such as quantum random access memories (qRAM) that allow classical information to be encoded in quantum mechanical form~\cite{arunachalam2015robustness}. A qRAM to access $N$ pieces of data consists of a branching array of $2N$ quantum switches, which must operate coherently during a memory call.   In principle, such a qRAM takes time $O(\log_2 N)$ to perform a memory call, and can tolerate error rates of up
to $O(1/\log_2N)$ per switching operation, where $\log_2 N$ is the depth of the qRAM circuit.   Proof of principle demonstrations of qRAM have been performed, but constructing large arrays of quantum switches is a difficult technological problem.  

These hardware challenges are technical in nature, and
clear paths exist towards overcoming them.   They must be
overcome, however, if quantum machine learning is to
become a `killer app' for quantum computers. As noted previously, most of the quantum algorithms that have been identified face a number of caveats that limits their applicability.  We can distill the caveats mentioned above into four fundamental problems.

\begin{enumerate}
\item Input problem: While quantum algorithms can provide dramatic speedups for processing data they seldom provide advantages for reading data.  This means that the cost of reading in the input can in some cases dominate the cost of quantum algorithms.  Understanding this potentially mitigating factor is a subject of ongoing work. 
\item Output problem: Learning the full solution from some quantum algorithms as a bit string requires learning an exponential number of bits.  This makes some applications of QML algorithms infeasible.  This problem can potentially be sidestepped by only learning summary statistics for the solution state but more work is needed.
\item Costing problem: Closely related to the input/output problem(s), at present very little is known about the actual number of gates required by quantum machine learning algorithms.  While bounds on the complexity suggest that for sufficiently large problems they will offer huge advantages, it is still an open question to determine exactly when that crossover point occurs.  
\item Benchmarking problem: It is often difficult to assert that a quantum algorithm is ever better than all known classical machine algorithms in practice because this requires extensive benchmarking against modern heuristic methods.  Additional results establishing lower bounds for quantum machine learning would partially address this.  
\end{enumerate}

One potential path forward that sidesteps some of these issues is examining applications of quantum computing to quantum, rather than classical, data.  The aim therein is to use quantum machine learning to characterize and control quantum computers~\cite{wiebe2015quantum}.   This would enable a virtuous cycle of innovation similar to that which occurred in classical computing, wherein each generation of processors is then leveraged to design the next generation processors.  We have already begun to see the first fruits of this cycle with classical machine learning being used improve quantum processor designs~\cite{sasaki2001quantum, 2010PhRvA..81c2324B, bisio2011quantumlearning, 2012arXiv1208.0663S, sentis2014quantum, paparo2014quantum, dunjko2015quantum-enhanced, dunjko2016quantumenhanced, sentis2016quantumchange,wan2016quantum,Laflamme2017,mavadia2017prediction} which in turn provide powerful computational resources for quantum enhanced machine learning applications themselves~\cite{2009PhRvL.103o0502H, 2013PhRvL.110y0504C, 2012PhRvL.109e0505W, lloyd2013quantum, 2014PhRvL.113m0503R, 2014arXiv1401.2142W, 2014NatPh..10..631L, lau2016quantum}.

\section*{Acknowledgments and author contributions section}

\noindent Financial acknowledgements.  J.B.~acknowledges AFOSR grant FA9550-16-1-0300, Models and Protocols for Quantum Distributed Computation, for financial support.   P.W.~acknowledges financial support from the ERC (Consolidator Grant QITBOX), Spanish Ministry of Economy and Competitiveness (Severo Ochoa Programme for Centres of Excellence in R\&D SEV-2015-0522 and QIBEQI FIS2016-80773-P), Generalitat de Catalunya (CERCA Programme and SGR 875), and Fundacio Privada Cellex.  P.R.~and S.L.~acknowledge funding from ARO and AFOSR under MURI programs.
\\ 

\noindent Origin of figures.  Figure~\ref{fig:aqc-gibbs} courtesy of Lusa Zheglova (illustrator).  Bar plots in Box 3 produced by the authors. \\

\noindent Author contributions.  All authors designed the study, analyzed data, interpreted data and wrote the article.  \\ 
  
\noindent Competing interests.  The authors declare no competing interests. \\


\end{document}